# COMPARISON OF SOLAR WIND SPEED ESTIMATES FROM NEARLY SIMULTANEOUS IPS OBSERVATIONS AT 327 AND 111 MHZ

*Chashei I.V.[1], Lukmanov V.R.[1], Tyul'bashev S.A.[1], Tokumaru M.[2]*

[1] - Lebedev Physical Institute, Russia

[2] - ISEE, Nagoya University, Japan

Corresponding author Chashei I.V., chashey@prao.ru

**Abstract**

Results are presented of a comparison between solar wind speed estimates made using the time delays between 3 pairs of 327 MHz antennas at ISEE and estimates made by modeling the temporal power spectra observed with the 111 MHz BSA antenna at LPI. The observations were made for 6 years in the descending phase of solar cycle 24. More than 100 individual records were obtained for the compact source 3C48 and the extended and anisotropic source 3C298. The correlation between the daily speed estimates from 3C48 is 50%. Their annual averages agree within the error estimates and show the expected solar cycle variation. However the correlation between speeds from 3C298 is only 25% and their annual averages do not agree well. We investigate possible causes of this bias in the 3C298 estimated speeds.

## 1. Introduction

Interplanetary scintillation (IPS) are widely used for investigation of the solar wind. Main advantage of this method is that it allows obtaining information about moving solar wind plasma in wide range of heliocentric distances and helio latitudes including ranges that are inaccessible now for local measurements. Solar wind speed can be measured if IPS are observed simultaneously at several spaced radio telescope (Hewish and Denisson, 1967; Vitkevich and Vlasov, 1966; Armstrong and Coles, 1972). At the moment, only ISEE has 3-site facilities for regular measuring of the solar wind speed; the spatial distribution of the solar wind speed and its evolution in the solar activity cycles were investigated during recent three decades (Kojima and Kakinuma, 1990; Tokumaru et al., 2012). Other possibility of solar wind speed estimates is based on the measurements of the IPS temporal power spectra (Manoharan and Ananthakrishnan, 1990). Comparison between 1-site and 3-site solar wind speed estimates (Manoharan and Ananthakrishnan, 1990; Glubokova et al, 2011; Mejia-Ambriz et al., 2015) showed sufficiently good agreement. In this paper we compare 1-site and 3-site speed estimates data during 6 yearlong observation series on the descending phase of 24 solar activity cycle. In contrast with (Manoharan and

Ananthakrishnan, 1990) comparing 1-site and 3-site speed estimates at one and the same frequency 327 MHz, our consideration is related to two strongly different frequencies, 111 MHz and 327 MHz.

The motivations of our study are the following. The mean velocity estimates from single station IPS can be sensitive to radio source size, the solar wind density spectrum, and, especially for low frequencies, to the distribution of turbulence along the line of sight. At the same time, high frequency multi-station measurements are based on the time delay between IPS at different locations, so they are less sensitive to effects such as the size of the radio source or the power spectrum of the solar wind density fluctuations. IPS velocity observations estimate the mean velocity over the line of sight. These can be tomographically inverted (Jackson B.V. et al, 1998) to provide a spatial map of the velocity for comparison with other solar observations. Future development of our analyses is directed to producing Carrington and daily velocity maps from BSA LPI data using, in particular, tomographic inversion of mean velocity observations. In this context, the combination of BSA LPI and ISEE data would be very useful for two reasons: i) BSA LPI is more sensitive than ISEE allowing observation of more rich population of scintillating sources and more detailed maps in adjacent solar wind regions; ii) velocity and scintillation level BSA LPI maps can be supplemented with ISEE data in the declination ranges higher than +42º and lower than -8º where BSA LPI data are not available.

**Observations and data analyses**

Below we compare between 3-site and 1-site data on solar wind speed for two strong scintillating radio sources 3C 48 with closest approach to the Sun in summer time and 3C 298 with closest approach in winter time.

1-site observations were conducted at the radiotelescope BSA LPI (Big Scanning Antenna of Lebedev Physical Institute) which is a beam array with 16384 half-wave dipoles. The geometrical sizes of the array is 200*400 m, the effective area is about 45000 m$^2$ towards zenith. After general reconstruction of the antenna finished in 2012 (Shishov, 2016) BSA had independent beam systems. One system has 128 fixed beams. 96 of those beams are connected to digital recorders and cover declinations from -8º to +42º. Since 2014 observations are conducted around the clock according to the program 'Space Weather'. The antenna is a meridian passage, so every radio source at the sky is observed during approximately 3.5 minutes by half-power at the polar pattern once a day. The central frequency of

BSA is 110.3 MHz with a total bandwidth of 2.5 MHz. The signal is recorded by a digital recorder at a frequency of 10 Hz. The typical sensitivity during observations of radio sources scintillating on the interplanetary plasma is 0.2 Jy.

Multi-site observations of IPS at 327 MHz have been conducted regularly since 1980s at ISEE, Nagoya University (Kojima and Kakinuma, 1990; Tokumaru, 2013). Cylindrical parabolic reflector antennas at Toyokawa, Fuji, and Kiso were employed for ISEE IPS observations in the analysis period (Tokumaru et al., 2011). The dimensions of Toyokawa, Fuji, and Kiso antennas are 108 m by 38 m, 20 m by 100 m, 27 m by 75 m, respectively. The phased array receivers are installed on these IPS antenna. The receiver bandwidth is set to be 10 MHz. The beam direction of the Toyokawa antenna is confined in the median and steerable between 60 degree south and 30 degree north with respect to the zenith, while those of Fuji and Kiso antennas are steerable in both north-south and east-south directions. Thirty to forty compact sources within solar elongation < 90 degree are observed in a day using three antennas simultaneously. IPS data are collected during about 3 minutes around the meridian transit at Toyokawa for a given source, and the sampling period is 20 ms. The solar wind speed $V_3$ is estimated using time delays of 3 temporal cross-correlation functions calculated for simultaneous IPS records at 3 spaced radio telescopes of ISEE (Tokumaru et al., 2012). The solar wind speed $V_1$ is estimated from temporal IPS power spectra measured at radio telescope BSA LPI with more low operating frequency 111 MHz. The fitting procedure of $V_1$ determination is described in details in the papers (Manoharan and Ananthakrishnan, 1990; Glubokova et al, 2011). We assume spherically symmetric spatial solar wind density distribution with constant radially directed speed and with the power exponent n = 3.6 of 3D spatial density turbulence spectrum, and the angular sizes $\theta_0 = 0.33$ arcsec for the source 3C 48 and $\theta_0 = 0.52$ arcsec for the source 3C 298 (Tyul'bashev et al., 2020). Following (Manoharan and Ananthakrishnan, 1990; Glubokova et al, 2011) we assume that the angular brightness distributions are isotropic Gaussian for both radio sources and the turbulence power spectrum is also isotropic. The speeds $V_1$ are found from the best fit of theoretical IPS power spectra with the speed as the only variable parameter into the measured spectra. The temporal spectrum of scintillations is defined by the formula:

$$P(f) = 4A\lambda^2 \int \frac{dz}{v_\perp(z)} \int dq_\perp \Phi_e(q) \sin^2\left(\frac{q^2 z'}{2k}\right) F^2\left(\frac{qz'}{k}\right)_{|q_\parallel = \frac{2\pi f}{v_\perp(z)}}, \quad (1)$$

where f is the time frequency, $\lambda$ is the radio wave length, $k = \frac{2\pi}{\lambda}$ is the wave number, the axis OZ is the direction along the sight line towards the observed source (fig. 1), z=0 corresponds to the P-point, $A = 5 \cdot 10^{-25}$ cm², $r_0 = \sin\varepsilon \cdot 1AU$, $v_\perp(z) = v\cos\varphi = v\frac{r_0}{\sqrt{z^2+r_0^2}}$ – projection of the solar wind speed on the image plane at the point (r, z), $\upsilon$ is the solar wind speed (v ≈ 400 km/s near the Earth), $q$ is the spatial frequency, $q_\parallel$ is the component of the spatial frequency along the view sight, $q_\perp$ is the component in the image plane perpendicular to the view sight, $q = \sqrt{q_\perp^2 + q_\parallel^2}$, $\Phi_e(q) = Cq^{-n}$ is the spatial spectrum of fluctuations of the electron density of the interplanetary plasma, n is the turbulence index, $F(q) = \left(\frac{1}{2\pi}\right)^2 \iint d^2\theta \exp(-ikq\theta) I(\theta)$ is the spatial spectrum of the observed radio source, $I(\theta)$ is the brightness distribution across the source, $z' = z + \cos\varepsilon \cdot 1AU$.

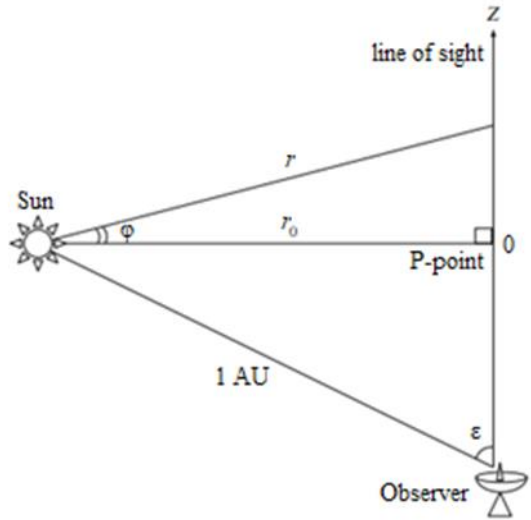

Fig. 1. Illustration for the observation of source scintillations at interplanetary plasma inhomogeneities (a case of spherical distribution of the plasma)

The data from the BSA of the Lebedev Physical Institute were processed as follows:

1) For every source and for every day of observation the theoretic moment when the source passes the radiation pattern peak was calculated;

2) The source record duration at BSA is $\frac{425^s}{\cos\delta}$, where δ is the declination of the radio source. For analysis the interval with the length twice as the source record duration and the center at the theoretic moment of the source pass through the pattern peak was chosen. The interval of this length is chosen to guarantee that it contains all the record of the radio source (the calculated and the real moments of the source pass through the pattern peak don't usually coincide due to the ionosphere influence or other reasons);

3) The source center was calculated (the real moment of the source pass through the pattern peak). To do this the considered interval was split into 32 smaller intervals of the same length about 30 sec ($\frac{425^s}{16\cos\delta}$ to be more exact). The signal was averaged at those 32 intervals to compensate noise and scintillation on one hand, and to save the information of the moment of the source pass through the pattern peak on the other hand. Then the convolution of the obtained array and the radiation pattern $\left(\frac{\sin x}{x}\right)^2$ at the interval (-π; π) was calculated: $s_i = \sum_{j=0}^{16} u_{i+j} \left(\frac{\sin x_j}{x_j}\right)^2$, where $u_k$ is the average signal at $k$-th interval ($k$=1..32). $x_j = -\pi + \frac{\pi}{8}j$, if $x_j \to 0$ then $\left(\frac{\sin x_j}{x_j}\right)^2 \to 1$. The convolution maximum $s_i$ was defined. To find the center more exactly the parabola passing points ($i - 1$; $s_{i-1}$), ($i$; $s_i$), ($i + 1$; $s_{i+1}$) was defined. The parabola vertex was considered as the source center;

4) The temporal scintillation spectrum was calculated using the Fast Fourier Transform of 2048 consecutive signal values with the source center in the midst. The obtained spectrum was split into 256 intervals each of those had 4 points. Each interval was averaged, thus the ultimate spectrum had 256 points;

5) Then the obtained spectrum was compared to theoretical spectra calculated by the formula (1) for the fixed values of the turbulence index (n = 3.6), angular size of the source and its elongation ε (angular distance between the directions to the Sun and to the source) closest to those at the observation moment. The comparison was made from 4[th] point of the spectrum to the point where the difference between the current spectrum value and the bottom noise value dropped by 30% compared to the difference between the top and bottom levels. The top level was defined as the average of values for points from 4[th] to 8[th] of the obtained spectrum. The bottom level was defined as the average of values for the last 20 points. The first point from the end for which the condition 'the values for this point and two closest left points exceed the level of *bottom_level + 0.7·(top_level – bottom_level)*' is true was chosen as the end point for comparison the obtained spectrum from observation and theoretical spectra. The end point was chosen this way because we can't define exactly the angular size of a radio source, so we cut the part which is influenced by the angular size a lot. We don't consider the first three points because the values for those are influenced by ionospheric scintillations. The best theoretical spectrum was chosen using the Least Squares Method (fig. 2). The speed for the chosen theoretical spectrum was considered to be the estimation of the solar wind speed at the moment of observation.

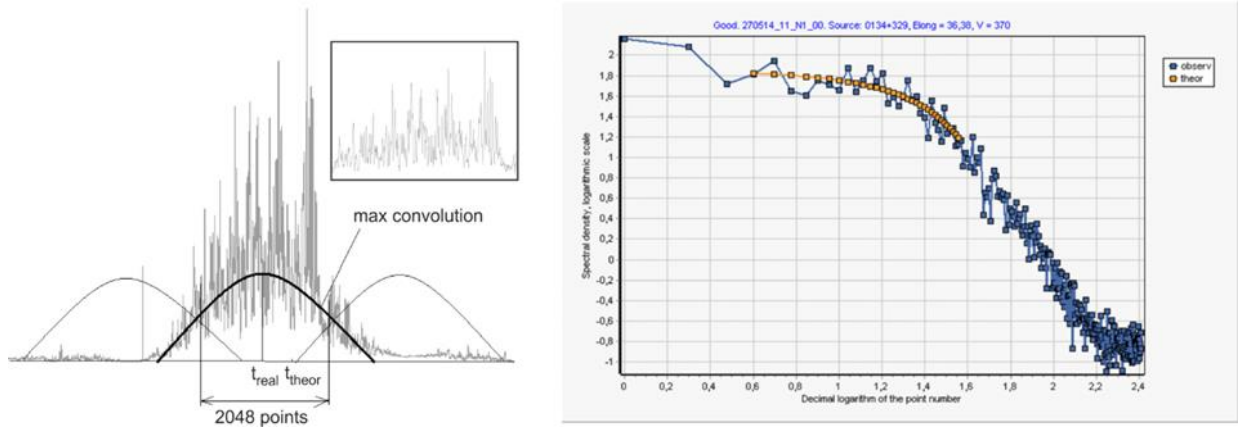

Fig. 2. An example of data processing. Left part is the initial record of 3C 48, ttheor and treal are theoretical right ascension (r.a.) calculated from coordinates of source and real r.a. from the convolution maximum. Fragment in the middle part represents 2048 central points used in calculation of the power spectrum. Right part is the observed temporal power spectrum of the source 3C 48 (blue) with the fitted theoretical spectrum (orange).

Theoretical spectra were calculated with assumptions similar to those in (Glubokova et. al, 2011, 2013):

1) the spatial spectrum of turbulence is power-mode: $\Phi_e(q) = Cq^{-n}$, where $C$ is so-called structure constant;

2) the structure constant depends on the distance from the Sun according to the power law: $C \sim r^{-4}$;

3) the solar wind speed is constant and the direction is radial;

4) the brightness distribution across the source is Gaussian, spherically symmetric: $I(\theta) = \exp\left(-\frac{\theta^2}{\theta_0^2}\right)$, where $\theta_0$ is the source radius at the level $\frac{1}{e}$ (Glubokova et al. (2013) defined $\theta_0$ as the radius at the level $\frac{1}{\sqrt{e}}$, so $I(\theta) = \exp\left(-\frac{\theta^2}{2\theta_0^2}\right)$, but in this paper we used the first definition). So, the spatial spectrum of the source will be $F(q) \sim \exp(-\frac{1}{2}\theta_0^2 k^2 q^2)$.

Theoretical spectra were calculated in advance for the value of the turbulence index n = 3.6, for the values of angular sizes from 0'' to 1'' with the step of 0.01'', for the values of elongations from 25° to 60° with the step of 1°, for the values of the solar wind speeds from 200 to 1200 km/s with the step of 10 km/s. The integration over z was made between the limits $z_{min} = -\cos\varepsilon \cdot 1 AU$ and $z_{max} = 2\,AU$ with the step of $\Delta z = 0.1\,AU$. Another $z_{max}$ values are discussed below. The integration over $q_\perp$ was made between the limits of 0 and $10^{-4}$ m$^{-1}$ with the step of $10^{-6}$ m$^{-1}$.

A comparison between the speed estimations obtained at BSA and corresponding estimations obtained at Institute for Space-Earth Environmental Research (ISEE) of Nagoya university in Japan has been made. ISEE estimating method is described in (Tokumaru et al., 2012). However, not all the speed

estimations were taken into consideration. We used the selection criteria described below:

1. Speed estimations were obtained for the days when the elongation of a given source ranged from $25^o$ to $60^o$. That's because at small elongations scintillation get to the saturated mode and are stifled by the source angular size. The accuracy of high frequency ISEE data is drop outside elongation of $60^o$. For this reason we have a lack of ISEE speed estimates for comparison. Besides that, at large elongations a model asymmetry is possible due to the influence of the near-earth region.

2. The scintillations were analyzed. To do this were made calculations of the dispersions at the peak of the radiation pattern and at its zero levels using the 3-point median filter with the step of 2 (among the points $1^{st}$, $3^{rd}$ and $5^{th}$ of the initial array the median point was considered in calculating the variance, then among points $2^{nd}$, $4^{th}$ and $6^{th}$ the median was chosen, then among point $3^{rd}$, $5^{th}$ and $7^{th}$ points was chosen median and so on). If the ratio of the variance in the peak to the smallest variance in one of the zero levels was less than 2, this day for this source was not considered.

3. If the number of the end point for comparison the obtained spectrum from observation and theoretical spectra described above was 12 or less, this day was not considered either. The narrowest temporal spectra were removed to exclude possible contribution from ionospheric scintillation.

4. The rest of the spectra were classified as 'good' and 'bad'. To do this the variance for each spectrum was calculated (root-mean-square deviation from the selected theoretical spectrum calculated from to $4^{th}$ point to the end point for comparison). Then for every source the average variance in each year was calculated. The spectra for which the variance was 1.5 times higher than the average variance were classified as 'bad', the rest were classified as 'good'. During the comparison of estimations from BSA to those from ISEE only 'good' spectra were considered.

This criterial is used to remove automatically from consideration the records which coincide with the calibration séances taking place at the radio telescope several times per day.

5. Among the rest points those which were too far from other point were also removed. To do this the approximation by a function $y = x + b$ was made, where x is the speed obtained at BSA, y is the speed obtained at ISEE. The value of b was defined using the Least Squares Method. Then the root-mean-square deviation of the points from the straight line. If the deviation of a certain point exceeded the root-mean-square deviation more than by the factor of 3, this point was removed from the consideration.

Criterial for signal to noise (S/N) ratio is attached to the end point of the spectrum. The power S/N is the ratio between low frequency IPS level and high frequency noise level. The condition at the end point "bottom_level + 0.7·(top_level – bottom_level)" corresponds to the relation N+0.7 (S-N) > BN, where B is boundary constant. One can find from this relation the boundary value S/N > (B-0.3)/0.7. In our simulations we assume the value B≈30 that gives S/N > 40. Typical values of S/N are S/N >500 for the source 3C 48 and S/N > 200 for the source 3C 298 at the elongations about 25º, that means that S/N boundary condition is fulfilled for almost all measured IPS power spectra.

The fractions of records removed by the above criteria are the following: 185-196 records during a year are not in the elongation limits (1); about 3 % (2); about 10 % (3); about 6% (4) and 100 – 140 records for 3C 48 and 70-120 records for 3C 298 during a year were not considered due to the lack of ISEE data. Initial data for 'good' spectra are available at site of Pushchino observatory (http://prao.ru/English/index.php → Online data)

The data obtained for 6 years from 2014 to 2019 related to the descending phase of the 24 solar activity cycles. All the IPS records with the source elongation angles between 25º and 60º are analyzed for each year; corresponding heliocentric distances of line of sight proximate point (P-point) are within the range 0.4 and 0.8 AU. The P-point helio-latitudes decrease from 50º to 10º with increase of solar elongation for each source.

Simultaneous IPS observations are carried out at three spaced ISEE radio telescopes Toyokawa, Fuji, and Kiso. The vector of the solar wind speed $V_3$ is estimated from the temporal shifts of three IPS cross-correlation functions maxima with known geometry of radio telescope locations. One can expect that the speed $V_3$ is less sensitive to the source size and to the model of solar wind plasma turbulence.

## 6. Estimates of the solar wind speed from the IPS power spectra

The data for solar wind speed IPS estimates are presented in the Figures 3 and 4 for the sources 3C 48 and 3C 298, respectively. ISEE speed $V_3$ and BSA LPI speed $V_1$ are shown on the vertical and horizontal axis's, the symbols for years from 2014 to 2019 are labelled on the panels. One can see from the Figures 3,4 that all speed estimates increase with decrease of the solar activity level. We calculated cross-correlation coefficient $\rho_c$ for the data on Fig.3, 4, they are equal to $\rho_{c48} = 0.48$ and $\rho_{c298} = 0.25$. Not very high cross-correlations mean that we have rather big velocity spread on the background of annual trend. The cross-correlation coefficients calculated over the year's interval are even lower than the values found for the whole interval. The spread is evidently connected with error noise in daily velocity estimates. One can see from the Table 1 that the velocity variances in BSA LPI and ISEE data in average are close each to other.

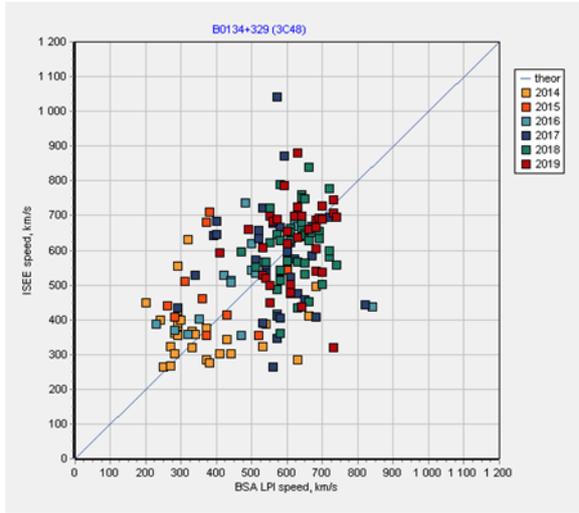

Fig. 3. ISEE solar wind speed vs BSA LPI solar wind speed for the source 3C 48, the symbols for the observation years are labelled on the panel, the diagonal corresponds to equality $V_1 = V_3$, the upper integration limit zmax in (1) is equal to 2 AU.

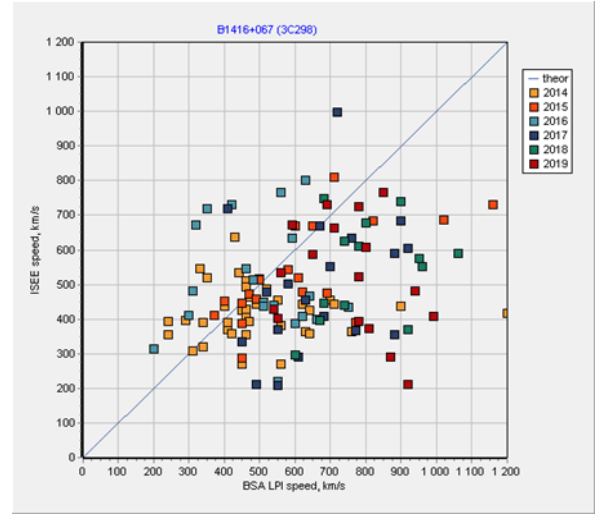

Fig. 4. The same as Figure 3 but for the source 3C 298.

The Figure 5, in which averaged over the years speeds are shown, also illustrates the tendency of the speed evolution in the activity cycle for both radio sources: the averaged speed increases from minimal values about 400 km/s for activity maximum in 2014 to more than 600 km/s for activity minimum in 2019. The diagonals of Figure 3 and 4 square panels is passing through the mid of data cloud that means that the speeds $V_1$ and $V_3$ in average are close each to other. Figure 5 also shows the approximate coincidence between average values $V_1$ and $V_3$. One can see from the Fig.5 that the BSA LPI speed estimates for the source 3C 48 are slightly lower than the ISEE speed estimates, but this relation is opposite for the source 3C 298. We can assume that the difference between $V_1$ estimates for compact radio source 3C48 and more extended radio source 3C 298 can be explained by not completely known model of modulating plasma distribution on the lines of sight.

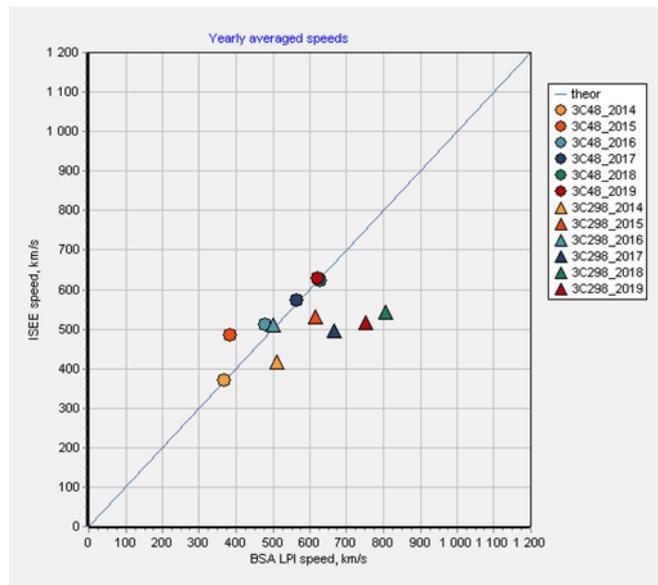

Figure 5 – Yearly averaged speeds for ISEE observations at vertical axis vs corresponding BSA LPI speeds at horizontal axis, the symbols are labelled on the panel.

To demonstrate it we calculated the difference $\Delta V = (V_1 - V_3)$ by modelling IPS power spectra with different upper integration limit in (1). The dependence of $\Delta V$ on upper integration limit $z_{max}$ is presented on the Fig. 6 for both radio sources in the range 1 AU $*\cos 45º < z_{max} < $ 2 AU. Fig. 6 shows that speed estimates are not sensitive to $z_{max}$, if $z_{max} \geq 2$ AU. The difference $\Delta V$ is very small, about several percent, in the case of compact source 3C 48 by $z_{max} = 2$ AU. In the case of more extended source 3C 298 the difference is minimal, about 10- 20 percent by integration in symmetrical limits – $\cos \varepsilon < z < \cos \varepsilon$. Such interpretation is confirmed by the calculation of model IPS power spectra with the enhanced density plasma slab located near the Earth orbit. The IPS power spectra for the model with the slab are presented in the Fig. 7a,b at the elongation 60º for the sources 3C 48 and 3C 298, respectively. The slab thickness is 0.2 AU, constant C (absolute density variance) in the slab is 4 times higher than in spherically symmetric model. We see from the Fig. 7 that introduction of the slab results in broadening of temporal spectra in comparison with spherically symmetric model. The broadening which is equivalent to increase in estimated speed $V_1$ is comparatively small for compact source, Fig 7a, and much more pronounced for extended source, Fig. 7b.

It should be noted that the speed data on the year 2015 deviates slightly from the general trend, at least for the source 3C 48. This peculiarity is somehow connected this the highest IPS level (averaged over the year scintillation index) in 24 solar cycle observed in 2015 (Tyul'bashev et al., 2020).

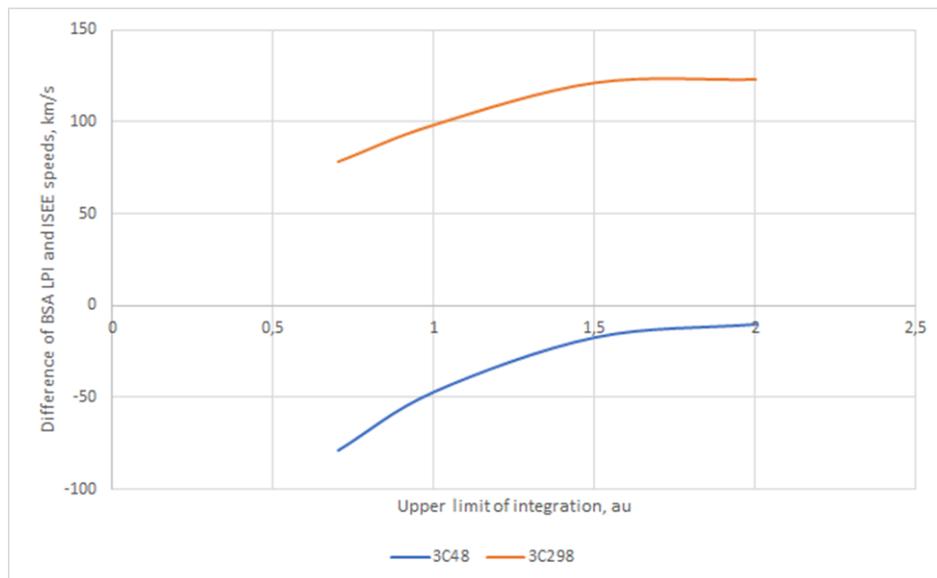

Figure 6 – The dependence of difference ($V_1$-$V_3$) on upper integration limit $z_{max}$.

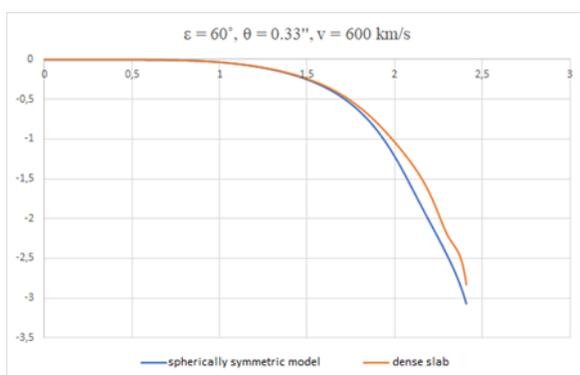

Figure 7a – Model IPS power spectra with the dense slab near the Earth, yellow, in comparison with the spherically symmetric model, blue, for the source 3C 48.

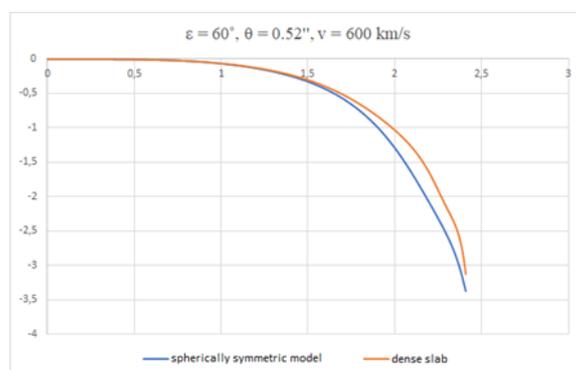

Figure 7b – The same as 7a, but for the source 3C 298.

Table 1 has additional information not included in fig. 1-5. The table shows yearly averaged speeds and mean-square deviations for sources 3C48 and 3C298. The table also includes number of points for each source in each year.

|  | 2014 | 2015 | 2016 | 2017 | 2018 | 2019 |
|---|---|---|---|---|---|---|
| BSA 3C 48 RMS annual variance | 368±24 127 | 385±29 96 | 479±32 137 | 565±15 98 | 629±9 58 | 623±13 76 |
| ISEE 3C 48 RMS annual variance | 369±16 84 | 485±34 113 | 511±27 114 | 571±22 139 | 623±15 98 | 627±19 112 |
| Number of points | 28 | 11 | 18 | 40 | 43 | 34 |
| BSA 3C 298 RMS annual variance | 510±29 184 | 615±44 199 | 500±31 140 | 667±35 152 | 806±37 134 | 754±33 136 |
| ISEE 3C 298 RMS annual variance | 417±12 73 | 530±30 132 | 512±35 155 | 496±44 192 | 543±38 138 | 517±38 158 |
| Number of points | 39 | 20 | 20 | 19 | 13 | 17 |

Data in Table 1 show that the yearly average BSA LPI and ISEE speeds are in close agreement for the source 3C 48, but the BSA LPI speed is systematically biased to higher velocities for the source 3C 298 excluding the year 2016. 3C 298 bias averaged over all data in Table 1 is about 140 km/s that is about 20 % from the mean speed, the value of bias increases with approach the solar minimum. .

## 7.   Discussion

The synchronous increase of the speeds $V_1$ and $V_3$ in the Figures 3-5 is in a good agreement with evolution of the solar wind global structure in the solar activity cycle. Indeed, in situ Ulysses measurements (McComas et al, 2003) and many years IPS observations (Tokumaru et al, 2012; Manoharan, 2012 ) have shown that the spatial solar wind structure is close to spherically symmetric in the period of solar maximum, and has the bi-modal character with low speed, 300-400 km/s, at small helio-latitudes and high speed, 700-800 km/s, at high helio-latitudes in the period of solar minimum. The region of transit from the slow wind to the fast wind is located at helio-latitudes about 20°. Bigger and bigger part of days from the year observation series has the lines of sight passing through the solar wind regions of the fast streams emanating from polar coronal holes when the solar activity level decreases, that explains the speed increase between 2014 and 2019 in the Figures 3-5.

As mentioned above (Manoharan and Ananthakrishnan, 1990; Glubokova et al, 2011; Mejia-Ambriz et al., 2015) have found reasonably good agreement between $V_1$ and $V_3$ estimates. Our study is based on more rich statistics that allow the comparison between 1-site and 3-sites speeds at maximal and descending phases of the solar activity cycle that can be considered as the base for using 1-site IPS measurements for study dynamics of the solar wind speed spatial distribution in the solar activity cycles.

Using different behavior of ΔV values presented in the Fig. 6 one can define the boundary between compact and extended sources. The angular size of the Fresnel scale from the formula (1) is equal to

$\theta_{Fr}$ = (2k 1AU cosε)$^{-1/2}$ . (2)

The value of $\theta_{Fr}$ is approximately equal to $\theta_{Fr}$ ≈ 0.3 arc sec for radio frequency 111 MHz at the mean elongation ε = 45°. Thus, the source with $\theta_0 \leq \theta_{Fr}$ (3C 48, $\theta_0$≈0.33 arc sec) can be classified as compact while the source with $\theta_0 > \theta_{Fr}$ (3C 298, $\theta_0$≈0.52 arc sec) can be considered as extended. The main contribution in radio wave modulation is produced by the region located in vicinity of P-point for the compact source and by the region located between the observer and P-point for the extended source. It follows from the Fig.6 that the best accuracy for the solar wind speed estimates from the IPS temporal power spectra will be achieved with the upper integration limit $z_{max}$ = 2 AU in the case of compact sources and with the symmetric integration limits ± cos ε in the case of extended sources. Using symmetric integration limits would be a way to reduce the $V_1$ speed bias for the extended source. The difference in the $V_1$ / $V_3$ relation between the extended source and compact source seems to be interesting finding of the above consideration. We think that the reason enhanced $V_1$ spread as well some systematic bias between $V_1$ and $V_3$ for extended source are caused by a deviation of modulating plasma distribution on the line of sight from spherically symmetric one. The layers located symmetrically relative to P-point produce comparable contribution in the IPS temporal power spectrum in the case of compact source, while the IPS level is more sensitive to the region located between P-point and the observer for the extended source, in particular to the solar wind plasma near the Earth orbit. An increase in the absolute turbulence level in close to the Earth region leads to an enrichment of IPS power spectrum by higher frequencies that results in increase in a visible speed $V_1$.

In above simulation of IPS power spectra we assumed that the turbulence is isotropic.

Chang et al. (2019), Chashei et al. (2000 a, b) consider more complicated turbulence model with anisotropic density irregularities by modelling theoretical temporal IPS power spectra. In particular, Chashei et al. (2000 b) have shown that turbulence anisotropy can result in difference in $V_1$ and $V_3$, $V_1 < V_3$. To check the influence of anisotropy on the results of $V_1$ estimates we include in our theoretical model the turbulence anisotropy assuming that density irregularities are elongated in radial direction with axial ratio 1.5. The can draw conclusion that the anisotropy of this kind does not influence practically on $V_1$ estimates. Indeed, the mean increase in $V_1$ for overall records is about 7 km/s for the source 3C 48 and about 10 km/s for the source 3C 298. Rather good correspondence between BSA LPI and ISEE data in isotropic model together with weak $V_1$ increase for 1.5 elongation show that the irregularities axial ratio does not differ strongly from 1 at the heliocentric distances about 0.5 AU, in contrast with considerable radial elongation in the range of the solar wind formation (Armstrong et al., 1990; Chashei et al., 2000a).

In order to investigate the influence of the model source diameter on $V_1$ estimate we calculated theoretical IPS power spectra for the source sizes 0.25 arc sec and 0.45 arc sec and compare the values $V_1$ with the value $V_1 = 370$ km/s found earlier for the source size 0.33 arc sec. The simulation results $V_1 = 350$ km/s for the size 0.25 arc sec and $V_1 = 420$ km/s for the size 0.45 arc sec show that possible changes in $V_1$ are less than 10 percent. The changes in the angular size for the anisotropic source 3C 48 in the range of elongation $25^o$ - $60^o$ according to our estimations are within the above angular sizes limits.

Another reason of the speeds bias is the influence of the source anisotropy. Modeling the IPS power spectra we assume simple one component symmetric model of the source 3C 298 angular structure. However the interferometric study (Fanti et al, 2002) shows that the shape of this source is more complicated. It consists from two nearly symmetric components of angular size 0.5 arc sec, the same as we used, separated by 1.6 arc sec in E-W direction. The observed bias between $V_1$ and $V_3$ can depend on the angle between the solar wind velocity vector and the major axis of the radio source. This angle changes during the passage of the source past the Sun. If the solar wind speed is directed perpendicular to the source major axis, the IPS level is defined by two approximately equal sources. The IPS of two sources will be uncorrelated because the angular separation between them is greater than the Fresnel angle. As a result, two IPS temporal spectra will be similar and the total IPS level decreases twice in comparison with the IPS level of the single source. If the solar wind speed is parallel to the source major axis the temporal power spectrum will contain low frequency component corresponding to the separation between the source components that will lead to increase in speed $V_1$ and, consequently, to increase in bias. We calculated the visible speed $V_1$ for the source with intermediate angular size 1 arc sec and found the increase in mean bias from about 140 km/s, see table 1 for the size 0.5 arc sec, to about 300 km/s. It should be noted that 3C 298 angular size 0.5 arc sec (Tyul'bashev et al, 2020) used in our study is based on analyses of scintillation index, this IPS characteristics, in contrast with temporal spectrum, is not sensitive to the source anisotropy due to 2D integration of the spatial spectrum. One can obtain the following spatial spectrum for the double source

$F^2(\mathbf{q}) = \exp(-q^2 z^2 \theta_0^2) \cos^2(z \mathbf{q} \theta_1)$ (3)

where $\theta_0 \approx 0.5$ arc sec are the angular sizes of two components and the vector $\theta_1$ corresponds to half of E-W angular separation between the centers of two components, $\theta_1 \approx 0.8$ arc sec. First factor in the relation (3) corresponds to the symmetric angular distribution of the sources brightness while the second factor describes the separation between two components. Using (3) in the relation (1) for temporal power spectrum we can find correct $V_1$ estimate, if the angle $\psi$ between solar wind speed and the source elongation is known from observation geometry. In particular, the velocity will be parallel to the source axis, $\psi=0$, for equatorial observations and perpendicular, $\psi = 90°$, for polar observations. This angle changes in the limits $23° < \psi < 90°$ in our observations. The low frequency knee in the temporal power spectrum is located in the frequency range 0.1 Hz – 0.2 Hz at small angle $\psi$ values. However, we involve in the analyses only higher frequencies, $f > 0.3$ Hz, in order to avoid possible contribution from ionospheric scintillation that, according to our observational experience with BSA LPI, can be significant just near the maximal elongations $\varepsilon \approx 60°$ and the smallest angles $\psi \approx 23°$. For this reason exact estimates of the speed $V_1$ using the anisotropic source model (3) for specific source 3C 298 is not possible at our operating frequency 111 MHz. Further work is needed to study the interesting problem of temporal IPS spectra for strongly anisotropic sources that can be realized or with sources having smaller angular separation between components or with observations at higher radio frequencies.

According to our estimations the density of scintillating sources on the sky is about 1 source per square degree, so BSA LPI can observe about several thousand sources daily. The recent survey at the close frequency 162 MHz (Chhetri R. et al. 2018) confirms this estimate and show that about 1/3 of these sources has the angular sizes 0.3 arc sec or even less. Thus, the number of sources like 3C 48 observed daily is about 1 000. As a prospect, we are going to involve the population of compact sources in obtaining mass unbiased speed estimates. The IPS study for more extended sources is also of interest because i) such sources are sensitive to the plasma close to the Earth orbit and ii) it will give the possibility for velocity bias corrections based on sources statistics.